\documentclass[aps,prl,superscriptaddress,footinbib,preprint]{revtex4-1} 
\usepackage{graphicx}
\usepackage{amsmath}
\usepackage{braket}
\usepackage{graphicx}
\usepackage{xcolor}  
\usepackage{siunitx}

\renewcommand{\vec}[1]{\boldsymbol{#1}} 

\newcommand{\RN}[1]{%
	\textup{\uppercase\expandafter{\romannumeral#1}}%
}

\newcommand{\mysize}{0.5}

\makeindex

\begin{document}
\title{Coherent Coupling of Remote Spin Ensembles via a Cavity Bus} 

\author{T. Astner}
\email{thomas.astner@tuwien.ac.at}
\affiliation{Vienna Center for Quantum Science and Technology, Atominstitut, TU Wien, Stadionallee 2, 1020 Vienna, Austria}
\author{S. Nevlacsil}
\affiliation{Vienna Center for Quantum Science and Technology, Atominstitut, TU Wien, Stadionallee 2, 1020 Vienna, Austria}
\author{N. Peterschofsky}
\affiliation{Vienna Center for Quantum Science and Technology, Atominstitut, TU Wien, Stadionallee 2, 1020 Vienna, Austria}
\author{A. Angerer}
\affiliation{Vienna Center for Quantum Science and Technology, Atominstitut, TU Wien, Stadionallee 2, 1020 Vienna, Austria}
\author{S. Rotter}
\affiliation{Institute for Theoretical Physics, TU Wien, Wiedner Hauptstraße 8-10/136, 1040 Vienna, Austria}
\author{S. Putz}
\affiliation{Vienna Center for Quantum Science and Technology, Atominstitut, TU Wien, Stadionallee 2, 1020 Vienna, Austria}
\affiliation{Department of Physics, Princeton University, Princeton, NJ 08544, USA}
\author{J. Schmiedmayer}
\affiliation{Vienna Center for Quantum Science and Technology, Atominstitut, TU Wien, Stadionallee 2, 1020 Vienna, Austria}
\author{J. Majer}
\email{johannes.majer@tuwien.ac.at}
\affiliation{Vienna Center for Quantum Science and Technology, Atominstitut, TU Wien, Stadionallee 2, 1020 Vienna, Austria}

\date{\today}
\label{par:abstract}
\begin{abstract} 
We report coherent coupling between two macroscopically separated nitrogen-vacancy electron spin ensembles in a cavity quantum electrodynamics system.
The coherent interaction between the distant ensembles is directly detected in the cavity transmission spectrum by observing bright and dark collective multiensemble states and an increase of the coupling strength to the cavity mode.
Additionally, in the dispersive limit we show transverse ensemble-ensemble coupling via virtual photons.

\end{abstract}

\maketitle
\label{par:intro}

The negatively charged nitrogen-vacancy (NV) center in diamond \cite{Jelezko2006} has attracted significant attention as it has long coherence times even at room temperature \cite{Jelezko2004} and has the possibility to act as transducer between the microwave and optical photon domain \cite{Togan2010,Blum2015a}.
Single NV centers have been successfully coupled over microscopic and macroscopic distances, using either direct dipole-dipole coupling \cite{Dolde2013} or spin-photon entanglement \cite{Bernien2013,Hensen2015}. In hybrid quantum systems \cite{Xiang2013a} strong coupling of different spin ensembles to a single mode cavity has been shown \cite{Kubo2010,Amsuss2011,Probst2013,Huebl2013,Ghirri2016}. 

In this letter we present an experiment that demonstrates coherent coupling between two spatially separated macroscopically distinct NV spin ensembles via a superconducting transmission line resonator.
For each of our ensembles we observe strong coupling to the cavity mode. Tuning the ensembles simultaneously into resonance we find collective bright and dark multi-ensemble dressed states and measure the discrete $ \sqrt{N} $ scaling of the dipolar coupling to the cavity mode. 
In the dispersive \cite{Brune1994} cavity quantum electrodynamics \cite{Mabuchi2002} regime we demonstrate transverse direct ensemble-ensemble coupling \cite{Sillanpaa2007,Majer2007} via virtual photons in the cavity. This opens the opportunity for the coherent quantum information transfer between remote solid-state spin ensembles.

\begin{figure}[t]
	\includegraphics[width=\mysize\columnwidth]{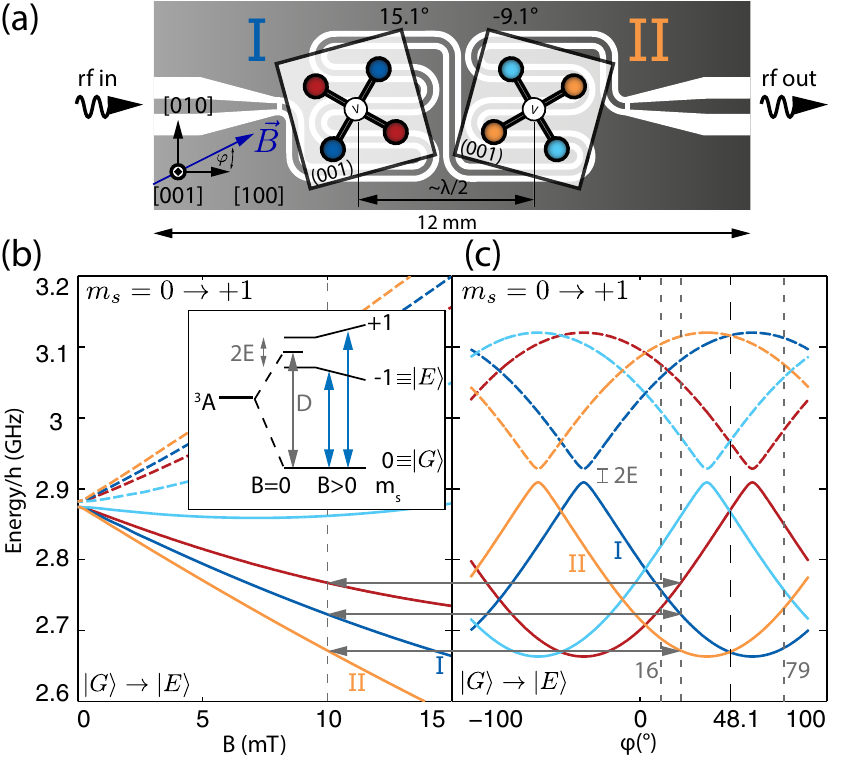}
	\caption{(Color online) \textbf{Experimental set-up.} (a) Transmission line resonator with spatially separated diamond crystals. In the (001) plane for each diamond two magnetically distinguishable ensembles can be identified (marked with color). (b) Calculated transition energies as function of the magnetic field amplitude for the field angle connected with arrows in (c). (c) Calculated transition energies as a function of the field angle in the (001) plane with field magnitude corresponding to the dashed grey line in (b). At \ang{48.1} (dashed black line) the transition energies of two distant ensembles, \RN{1} and \RN{2}, are degenerate. Anti-crossings in this illustration correspond to crystal strain fields $ E $ which mix the $ m_s = -1 $ and the $ m_s = +1 $ state.}
	\label{fig:setup_tuning}
\end{figure}

Our experimental set-up is composed of two spatially separated diamond crystals containing NV defect centers, bonded individually onto a superconducting resonator (the distance between the crystals is approximately \SI{5}{\milli\meter}). The sample is surrounded by a 3D Helmholtz coil providing D.C.\  magnetic fields in arbitrary directions. The $ \lambda $ coplanar waveguide transmission line resonator used for read out and coupling of the two distant ensembles is made out of a niobium thin film on a sapphire substrate using optical lithography. In contrast to previous experiments \cite{Amsuss2011} we use the second resonance, i.e.\ the first harmonic $ \lambda $ resonance, which is found at $ \omega_c/ 2\pi =  \SI{2.7491}{\giga\hertz}$ with a quality factor of {$ Q \approx $ \SI{4300}{}} (limited by the surface losses of the crystals).
This $ \lambda $ resonance has two antinodes of the magnetic field at positions $ \lambda/4 $ and $ 3 \lambda/4 $ where the two crystals are positioned. The transmission line resonator depicted in Fig.~\ref{fig:setup_tuning} (a) is structured with several turns in order to maximize the coupling of the individual diamond crystals to the cavity mode. 

We use two type-Ib high pressure high temperature (HPHT) diamond samples with a concentration of ${ \approx   \SI{6}{ppm}}$ of NV centers \cite{Nobauer2013} each. The electronic ground state spin $ S=1 $ triplet can be described with a Hamiltonian \cite{Jelezko2006} of the form
\begin{align}
	H_{NV} = \hbar D S_z^2 + \hbar E(S_x^2 - S_y^2) + \hbar g \mu_b \vec{B_0}\vec{S},
\end{align}
with a zero-field splitting $ D/2\pi \approx \SI{2.87}{\giga\hertz}$ for the axial component along the NV axis and a strain field splitting $ E/2\pi \approx \SI{13}{\mega\hertz}$ in the transverse direction \cite{Wrachtrup2006}. The third term describes the interaction with an external static magnetic field ($ \vec{B_0} $) for Zeeman tuning of the $ m_s = \pm 1 $ states. We apply a field in the crystal (001) plane (parallel to the resonator plane) to lift the $ m_s = \pm 1 $ degeneracy and tune the transition energies. In the experiment we only use one $ m_s = 0 $ to $ m_s = -1 $ transition of each crystal, labeled with \RN{1} and \RN{2} in Fig.~\ref{fig:setup_tuning}(a).

As shown in Fig.~\ref{fig:setup_tuning}(a), we position each sample such that the relative angle between them is \ang{24.2}. The projection of the external magnetic field onto the NV axis is different for each sample which allows individual control of the transition energies of ensemble \RN{1} and \RN{2}. In Fig.~\ref{fig:setup_tuning}(b) and (c) we present the dependence of the $ m_s = 0 \rightarrow m_s = \pm 1 $ transition energies on the angle and magnitude of the external magnetic field.
In the experiment we perform transmission spectroscopy of the cavity and determine the $ S_{21} $ transmission scattering amplitude with a vector network analyzer.
To ensure thermal polarization in the ground state well above \SI{80}{\percent}, all experiments are carried out at \SI{60}{\milli\kelvin}.

We model the system using the Tavis-Cummings Hamiltonian \cite{Tavis1968} extended to two ensembles of $ N_{\RN{1},\RN{2}} $ spins each. Since cooperative collective effects play the major role in this experiment, it is instructive to introduce a Hamiltonian with collective spin operators
\begin{align}
	\begin{split}
		H_{\mathrm{eff}} &=\hbar\omega_ca^{\dagger}a + \hbar \omega_\RN{1} J_\RN{1}^z + \hbar \omega_\RN{2} J_\RN{2}^z + \\
		&+ \hbar g_\RN{1} (a J_\RN{1}^+ + a^\dagger J_\RN{1}^-) - \hbar g_\RN{2} (a J_\RN{2}^+ + a^\dagger J_\RN{2}^-).
	\end{split}
	\label{eq:hamiltonian}
\end{align}
Here $ a^\dagger a $ is the cavity photon number operator while the second and third part corresponds to the transition energy of each ensemble. The last two terms describe the interaction of each spin ensemble with the cavity mode. The collective operators are defined by $ J_{\RN{1}}^z = 1/2 \sum_{j=1}^{N} \sigma_{\RN{1}}^z $ and $ J_{\RN{1}}^\pm = 1/g_\RN{1} \sum_{j=1}^{N} g_j \sigma_{\RN{1}}^{\pm} $. The coupling rate scales with $ \sqrt{N} $ \cite{Dicke1954} and is given by $ g_\RN{1} = \sqrt{\sum\limits_{j=1}^{N_\RN{1}} |g_j|^2} $.
Operators for the second ensemble ($ \RN{2} $) are similarly defined.
As mentioned the $ \lambda $ cavity provides two anti-nodes of the magnetic field amplitude which have inverted sign. We account for this by the negative sign of $ g_\RN{2} $ in the Hamiltonian (Eq.~\ref{eq:hamiltonian}). 

In our first set of measurements we perform transmission spectroscopy of the system as a function of the magnetic field amplitude for constant angles in the (001) plane. With the field angle set to \ang{79}(\ang{23}) we tune the central spin frequency of the ensemble \RN{1}(\RN{2}) into resonance with the cavity mode.
We observe an avoided crossing for each ensemble with the corresponding polariton modes of the coupled cavity spin system \cite{Kubo2010,Amsuss2011}. From the measured $ S_{21} $ scattering amplitude at the avoided crossing we determine the collective coupling rates as $ g_\RN{1}/2\pi = \num{7.5} \pm \SI{0.1}{\mega\hertz}$ and $ g_\RN{2}/2\pi = \num{5.6} \pm \SI{0.1}{\mega\hertz}$. The polariton modes' decay rates are $ \Gamma_\RN{1}/2\pi = \num{2.45} \pm \SI{0.18}{\mega\hertz}$ and  $ \Gamma_\RN{2}/2\pi = \num{2.28} \pm \SI{0.16}{\mega\hertz}$ (half-width at half-maximum). We have chosen crystals with different parameters to explicitly distinguish them \cite{Nobauer2013}.

\begin{figure}
	\includegraphics[width=\mysize\columnwidth]{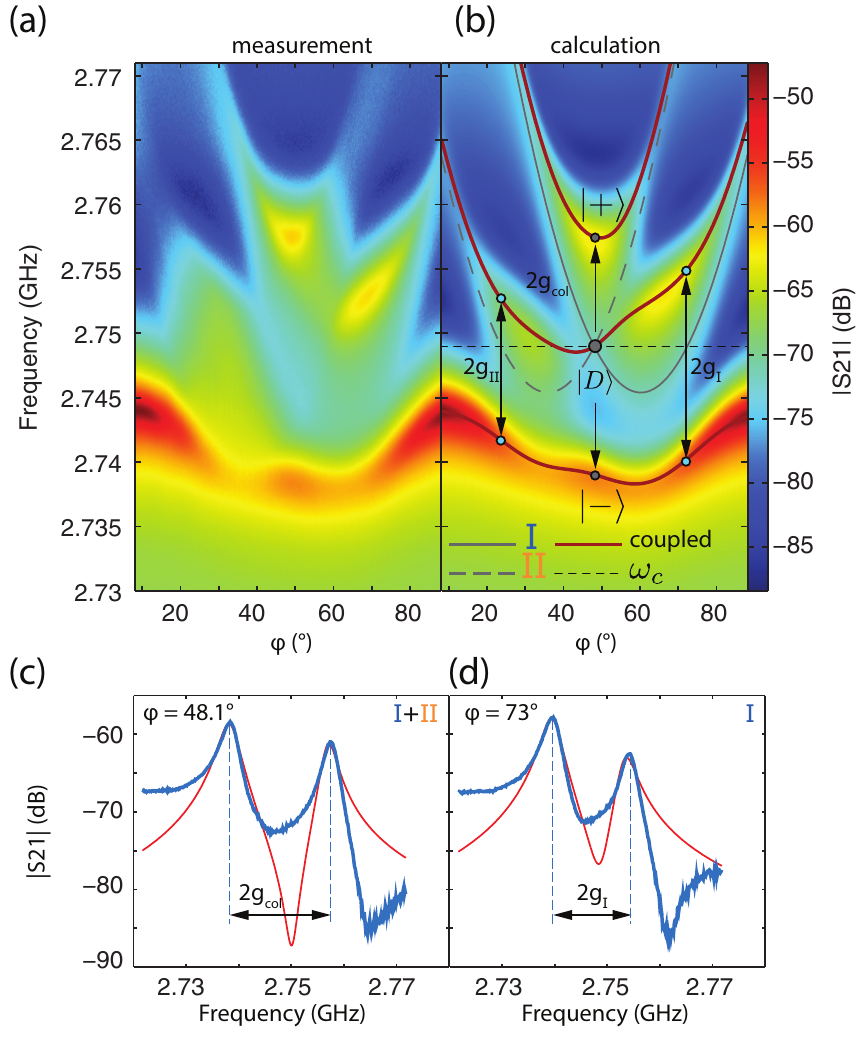}
	\caption{(Color online) \textbf{Transmission spectroscopy.} (a) Measurement data of the angle resolved transmission spectroscopy. At an angle of \ang{48.1} both ensembles coherently couple with the cavity mode and the coupling rate is enhanced compared to the individual coupling rates. Between the polariton modes $ \ket{\pm} $ a dark state $ \ket{D} $ emerges. (b) Calculated transmission spectrum with unperturbed eigenvalues in gray. The solid line corresponds to crystal \RN{1} and the dashed line to crystal \RN{2}. The horizontal dashed black line shows the unperturbed cavity resonance frequency. The splitting of the polariton modes is labeled with $ 2g_\RN{1} $ and $ 2g_\RN{2} $. At \ang{48.1} the resulting collective splitting of the system is labeled with $ 2g_{col} $. (c-d) Transmission spectroscopy measurement of the degenerate system (c) compared to a single ensemble (d).}
	\label{fig:spec_results}
\end{figure}

Consecutively, we perform angle resolved transmission spectroscopy by rotating the magnetic field in the (001) plane with a fixed magnitude.
The results shown in Fig.~\ref{fig:spec_results}(a) display three distinguishable features: At an angle of \ang{23} and \ang{79} the individual ensembles are in resonance with the cavity and we observe an avoided crossing. Moreover, at an angle of \ang{48.1} both ensembles and the cavity are degenerate. This results in an avoided level crossing which is much more pronounced and reflects the coherent coupling of all three constituents. From the transmission data, we derive the collective coupling strength $g_{col}/2\pi = \num{9.6} \pm \SI{0.1}{\mega\hertz}$ and decay rate of the collective polariton modes $ \Gamma_{\mathrm{col}}/2\pi = \num{1.49} \pm \SI{0.07}{\mega\hertz}$. At this point, the distant ensembles behave like an effective single ensemble with the predicted coupling rate $ \sqrt{g_1^2 + g_2^2}/2\pi \approx  \SI{9.36}{\mega\hertz}$.

We calculate the eigenenergies by diagonalizing the Hamiltonian of the coupled system Eq.~\ref{eq:hamiltonian}, shown as solid red lines in Fig.~\ref{fig:spec_results}(b). By comparing the measurement to these eigenvalues we confirm that at \ang{48.1} the system hybridizes into two polariton modes, an antisymmetric superposition of two Dicke \cite{Dicke1954} states of the distant ensembles and the cavity mode. A single excitation is thus shared between the two ensembles and the cavity. In the calculated transmission spectrum shown in Fig.~\ref{fig:spec_results}(b) we labeled the eigenstates with
\begin{align}
	\begin{split}
	\ket{\pm} &= \frac{1}{\sqrt{2}g_{\mathrm{col}}} [ \pm g_{\mathrm{col}} \ket{G^{\RN{1}}G^{\RN{2}}}_s \ket{1}_c - \\
	&- (g_\RN{1} \ket{ E^{\RN{1}} G^{\RN{2}}}_s - g_\RN{2} \ket{ G^{\RN{1}} E^{\RN{2}}}_s) \ket{0}_c]
	\end{split}
\end{align}
Here we introduce the excited state $ \ket{E^\RN{1}} = 1/N \sum_{j \in \RN{1}} \ket{g...e_j...g} $ in the form of a Dicke state for the first ensemble and similarly for the second. Next $ \ket{G} $ refers to the ground state of the ensembles.
The third state lies between the polariton modes and is a symmetric state in the form:
\begin{align}
	\ket{D} = \frac{1}{g_{\mathrm{col}}}\left(g_\RN{2}\ket{E^{\RN{1}}G^{\RN{2}}}_s + g_\RN{1} \ket{G^{\RN{1}} E^{\RN{2}}}_s\right) \ket{0}_c.
\end{align}
This state is not observable in transmission spectroscopy measurements, as it is a dark state of the system. 

Using the input-output formalism \cite{Walls2008} we calculate the transmission spectrum as shown in Fig.~\ref{fig:spec_results}(b) using the parameters of the individual ensembles such as coupling strength and line-width as well as the line-width of the cavity. The calculation shows good qualitative agreement with the measured data, including the formation of the dark state and increased coupling strength of the degenerate system. 

\begin{figure}[ht]
	\includegraphics[width=\mysize\columnwidth]{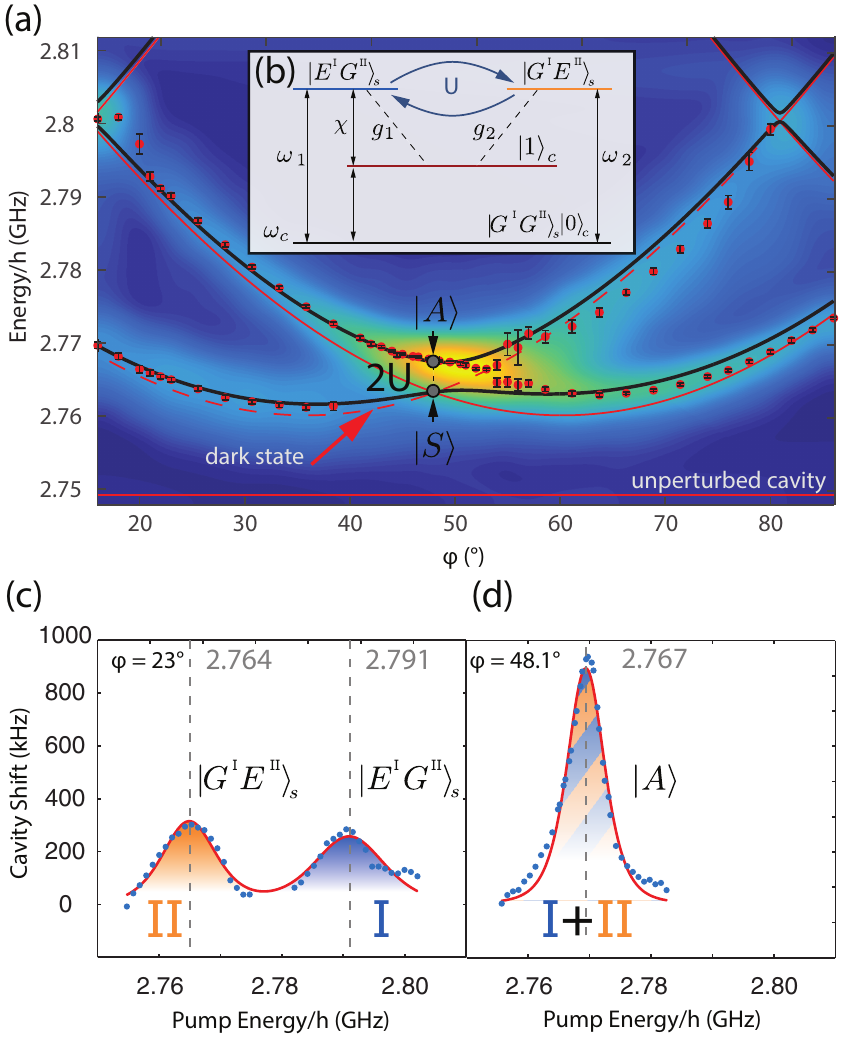}
	\caption{(Color online) \textbf{Dispersive level spectroscopy.} (a) Dispersive transmission spectrum with coupled eigenvalues in black solid lines and uncoupled eigenvalues of ensemble \RN{1}(\RN{2}) in red solid (dashed) lines. Data points mark the peak positions of the fitted resonances. In the vicinity of \ang{48.1} only the bright state of the coupled system is visible. (b) Ensemble-ensemble coupling scheme. The detuned spin ensembles produce a shift $ \chi $ on the cavity resonance. The ensembles interact with each other via the exchange of virtual photons in the cavity with the coupling rate $ U $. (c-d) Measured cavity shift for \ang{23} with the individual spin transitions visible. At an angle of \ang{48.1} only one prominent resonance is observed, corresponding to the antisymmetric state of the coupled system.}
	\label{fig:dispersive}
\end{figure}

So far we have shown coherent coupling of the two distant ensembles to the cavity mode.
In an experiment in the dispersive regime we now show direct transverse coupling between the distant spin ensembles without populating the cavity mode. To enter the dispersive regime we tune the magnitude of the external magnetic field such that the spin transitions and the cavity have an energy mismatch of $ \Delta_{\RN{1},\RN{2}} = \omega_c - \omega_{\RN{1},\RN{2}} > g_{\mathrm{col}} $. In all our dispersive measurements we maintain $ \Delta_{\RN{1},\RN{2}} \ge  \SI{12}{\mega\hertz}$. Despite this energy mismatch an interaction between the distant spin ensembles is still present due to the exchange of virtual photons via the cavity \cite{Blais2007,Filipp2009}. This coupling scheme is illustrated in Fig.~\ref{fig:dispersive}(b).
The dispersive interaction of the spin ensembles with the cavity mode leads to a frequency shift $ \chi_{\RN{1},\RN{2}} = g_{\RN{1},\RN{2}}^2/\Delta_{\RN{1},\RN{2}} $ of the cavity resonance. Within second order perturbation theory it is proportional to the coupling strength $ g_{\RN{1},\RN{2}} $ and the detuning $ \Delta_{\RN{1},\RN{2}} $.
With a drive signal resonant with the spin transitions, but off resonant to the cavity mode, a fraction of the ensemble is brought in a statistical mixture. Sequentially, the spin state dependent cavity shift $ \chi \braket{S_z} $ is monitored by weakly probing the cavity resonance.

In Fig.~\ref{fig:dispersive}(a) we show the measured cavity shift as a function of the pump frequency, while rotating the magnetic field to tune both ensembles through resonance. At an angle of \ang{23} the individual spin transitions are distinguishable and can be mapped to ensemble I and II (see Fig.~\ref{fig:dispersive}(c)).
By rotating the magnetic field to the degeneracy point at \ang{48.1} we observe a single prominent feature (see Fig.~\ref{fig:dispersive}(d)).
This resonance corresponds to the bright state of the coupled ensemble-ensemble system, which is an anti-symmetric superposition of an excitation in each of the ensembles. 
Furthermore, the coupling manifests itself in the appearance of a dark state in the vicinity of the degeneracy point.

We can model this measurement by using the Hamiltonian of (\ref{eq:hamiltonian}) and moving to the dispersive regime parameters with an effective Hamiltonian \cite{Blais2004a,Blais2007}:
\begin{align} 
	\begin{split}
		H_U &= \hbar(\omega_c + \chi_\RN{1} J_\RN{1}^z + \chi_\RN{2} J_\RN{2}^z)a^\dagger a \\
		&+\frac{\hbar}{2}[(\omega_\RN{1}+\chi_\RN{1}) J_\RN{1}^z + (\omega_\RN{2} + \chi_\RN{2})J_\RN{2}^z] \\
		&+ \hbar U(J_\RN{1}^- J_\RN{2}^+ + J_\RN{2}^- J_\RN{1}^+).
	\end{split}
	\label{eq:dispHam}
\end{align}
Here the first term describes the cavity mode which is shifted by $ \chi_{\RN{1},\RN{2}} $, while the second part denotes the spin transitions comprising the Lamb shift $ \chi_{\RN{1},\RN{2}} $ due to the presence of virtual photons \cite{Fragner2008}.
An effective transverse coupling of the remote ensembles is mediated via virtual photons. The coupling rate is given by $ { U = \frac{g_\RN{1} g_\RN{2}}{2} (1/\Delta_\RN{1} + 1/\Delta_\RN{2}) } $ and is described by the third part of the Hamiltonian (Eq.~\ref{eq:dispHam}).
The measured bright state at the degeneracy point can be identified as the antisymmetric state $ \ket{A} $ and the dark state as the symmetric state $ \ket{S} $. These states have the following form:
\begin{align}
	\ket{A} &= \frac{1}{g_{\mathrm{col}}} \left( g_\RN{2} \ket{G^{\RN{1}} E^{\RN{2}}}_s - g_\RN{1} \ket{E^{\RN{1}} G^{\RN{2}}}_s\right), \\
	\ket{S} &= \frac{1}{g_{\mathrm{col}}} \left( g_\RN{1}\ket{G^{\RN{1}} E^{\RN{2}}}_s + g_\RN{2}  \ket{E^{\RN{1}} G^{\RN{2}}}_s\right).
\end{align}
Since the ensembles are separated by a distance of $ \lambda/2 $, the A.C.\ magnetic field amplitude in the cavity has opposite sign for each ensemble. This results in an antisymmetric drive, which does not allow to drive any transition to the symmetric state \cite{Gywat2006,Blais2007,Filipp2011} and therefore this state is dark.
From calculations we infer that the coupling between the ensembles is given by ${ U/2\pi = g_\RN{1} g_\RN{2} /2\pi\Delta \approx \SI{2.2}{\mega\hertz} }$, where $ \Delta $ is the detuning of both ensembles to the cavity at \ang{48.1}. 

In conclusion, we have shown the coupling of two macroscopically separated spin ensembles coherently interacting with each other via the cavity.
Symmetric coupling between Dicke states of the distant ensembles leads to a dark state that forms between the polariton modes. The coherent collective coupling to the cavity mode is enhanced by a factor $ \sqrt{2} $, demonstrating that both ensembles behave like a single ensemble with twice as many spins.
In a dispersive level spectroscopy we show the existence of a transverse ensemble-ensemble coupling via virtual photons in the cavity. Furthermore, we observe a bright and a dark state of the dispersively coupled spin ensembles, which shows the coherent coupling between them.

Our system offers the flexibility to couple and decouple all individual elements in-situ, allowing to realize many different configurations. It should be noted that coherent coupling is necessary but not sufficient for observing entanglement in our system. It remains an interesting future challenge, both theoretically \cite{Liu2016} and experimentally, to create and detect entanglement in a system of spins coupled to a cavity.
Furthermore, our architecture is not limited to two ensembles and naturally offers the possibility to be extended to several ensembles coupled to each other while maintaining full control over the coupling between all elements of the system \cite{Zou2014a}. This shows the potential of ensembles of NV spins coupled to transmission line cavities for implementation of quantum information networks.

We thank W.J. Munro, M. Trupke, D. Krimer and P.Strassmann for fruitful discussions. The experimental effort has been supported by the TOP grant of TU Wien. T.A., A.A. and S.P. acknowledge the support by the Austrian Science Fund (FWF) in the framework of the Doctoral School ``Building Solids for Function" (Project W1243).

\bibliography{library}		

\end{document}